\documentclass{article}

\usepackage{PRIMEarxiv}
\usepackage[section]{placeins}
\usepackage[utf8]{inputenc} % allow utf-8 input
\usepackage[T1]{fontenc}    % use 8-bit T1 fonts
\usepackage{hyperref}       % hyperlinks
\usepackage{url}            % simple URL typesetting
\usepackage{booktabs}       % professional-quality tables
\usepackage{amsfonts}       % blackboard math symbols
\usepackage{nicefrac}       % compact symbols for 1/2, etc.
\usepackage{microtype}      % microtypography
\usepackage{lipsum}
\usepackage{fancyhdr}       % header
\usepackage{graphicx}       % graphics
\graphicspath{{./media/}}     % organize your images and other figures under media/ folder

\title{SerPyTor: A distributed context-aware computational graph execution framework for durable execution}
\author{
  Anuran Roy \\
  School of Computer Science and Engineering \\
  Vellore Institute of Technology \\
  Vellore, India\\
  \texttt{anuran.roy2020@vitstudent.ac.in} \\
  \And
  Sridhar Raj S \\
  School of Computer Science and Engineering \\
  Vellore Institute of Technology \\
  Vellore, India\\
  \texttt{sridharraj.s@vit.ac.in} \\
}

\begin{document}
\maketitle

\begin{abstract}
	% \lipsum[1]
	Distributed computation is always a tricky topic to deal with, especially in context of various requirements in various scenarios. A popular solution is to use Apache Spark with a setup of multiple systems forming a cluster. However, the prerequisite setup for a Spark cluster often induces an additional overhead, often limiting usage in constrained scenarios, especially in scenarios requiring context propagation. In this paper, we explore a relatively lightweight computational graph execution framework requiring little setup and fast speeds, coupled with context awareness.
\end{abstract}

% keywords can be removed
% \keywords{First keyword \and Second keyword \and More}
\keywords{Computational Graphs \and Distributed Computing \and Durable Execution \and Python}

\section{Introduction}
\label{sec:intro}

Distributed computing is a crucial aspect of services distributed across the internet. Most operations of services that we consume on an everyday basis, from ML recommendation algorithms on deciding the user's next choice at scale, to developer platforms helping developers deploy code faster. As such, setting up distributed compute clusters is a basic requirement in almost all facets across web services with varying degrees of use, but targeting a similar objective - reducing computation latency for large-scale computations. Tools like Apache Spark are available, which are language-agnostic in nature, but require proper setup and tuning.

In this paper, we introduce a more compact framework for performant execution of distributed computational graphs.

We discuss the physical hardware layer abstraction architecture and the logical execution flow in separate forthcoming sections bearing the same name, that are sections~\ref{sec:physical_abstraction} and \ref{sec:exec_flow} respectively.

\section{Literature Review}
\label{sec:lit_review}

% \lipsum[4] See Section \ref{sec:lit_review}.
According to \textit{Brezillion et. al.}\cite{brezillon1999context}, context is defined as the collection of relevant conditions and surrounding influences that make a situation unique and comprehensible. Further understanding and definition of context can be found on the same paper. In a subsequent paper, \textit{Pomero and Brezillon et. al}\cite{pomerol2001some} showed that there is a strong correlation between context and knowledge, and knowledge has been divided into three types of categories in their 1999 paper\cite{brezillon1999context}: \textbf{external knowledge}, \textbf{contextual knowledge} and \textbf{proceduralized context}.

The culmination of context-based policy works resulted in the formulation of the conceptual \textit{context-aware framework by Brezillon et. al}\cite{brezillon2017elaboration}. Graphs are mentioned to be one of the data structures that fit the context usage scenarios really well, according to \textit{Baldan, Paolo, et. al}\cite{baldan2000modelling}. Similarly, graph-based systems can act as vertex-centric models\cite{liu2020large}, adopting for more context usage and relevance, while passing data to its child node(s) in a discrete, well-defined (not necessarily well-structured) manner.

\section{Physical Layer Abstraction Architecture}
\label{sec:physical_abstraction}

\begin{figure} [!hb]
	\centering
	\includegraphics[width=18cm, height=8cm]{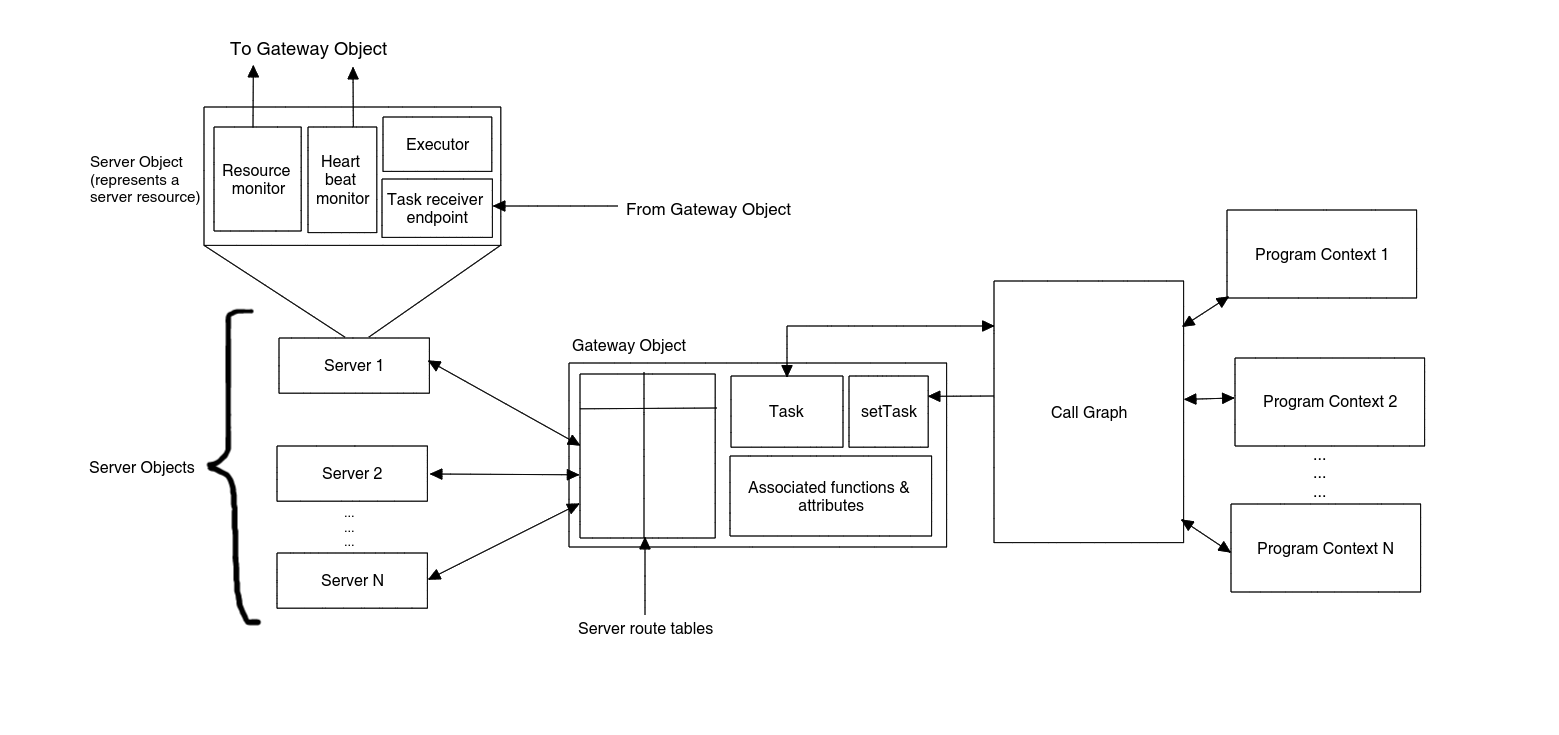}
	\caption{Hardware Abstraction Architecture}
	\label{fig:graph_arch}
\end{figure}

The Physical Layer Abstraction in SerPyTor comprises of primarily 3 objects: \textbf{HeartbeatServer}, \textbf{Server}, and \textbf{Gateway}, as shown in Figure \ref{fig:graph_arch}

The latter two can be customized to be fully generic as per the requirements, while the first object (that is, the \textbf{HeartbeatServer}) has a few limitations, primarily because it is a component of the generic \textit{Server} class, and has a few core functionalities to perform.

\subsection{HeartbeatServer}
\label{subsec:heartbeat_server}
The \textit{\textbf{HeartbeatServer}} object basically functions as the resource monitor for a server, and a successful response from it indicates that the server hardware is up and running. It reports in the form of a JSON response that reports the different types of resource usage for the server resource - for example, CPU usage, disk usage, (possible) GPU usage and memory usage, which indicates the status of the server to take the next task.

\subsection{Server}
\label{subsec:server}
The \textit{\textbf{Server}} is the abstract representation of a connected server in a network which is assigned to do computational tasks that require computational power / storage that the client's PC cannot perform (thus, the main motivation of the framework). The Server object exposes one or more ports from the physical server resource and responds to the request by either performing the computation or refusing it. The task to ascertain whether the server can take the task is delegated to the \textit{Gateway} object, so that the task to determine the optimal computational resource is delegated to a central authoritative entity to reduce conflicts at high concurrency. As such, the task of the gateway to determine optimal resources should be successfully executed as fast as possible.

The Server object is generic and weakly opinionated in nature. Every component in it can be modified as per the user's requirements - including mappings, execution mechanism, etc. Users can extend it with various user-defined or pluggable mechanisms - security check pipelines, authentication and authorization mechanisms, etc. Minimal number of assumptions have been made to impose minimal constraints on the user and to provide maximum programming flexibility.

The basic assumptions made in case of constructing a Server object are:

\begin{enumerate}
	\item The Heartbeat Server is a separate process than the Application Server, and thus, requires a separate port.
	\item Each mapping is a function that gets all its dependencies through Dependency Injection mechanism, so that the function itself forms an atomic task for durable execution (mentioned in Section \ref{subsec:durable_execution}).
\end{enumerate}

Separating the HeartbeatServer and Server processes, though might seem to be uncalled for, actually would help users, and additionally programs to troubleshoot and distinguish between system-level or application-level errors.

\begin{itemize}
	\item If there is a System-level error, then the corresponding Server Application would fail automatically.
	\item If there is an Application-level error, then the HeartbeatServer object would respond with a successful response. On the other hand, the Server object would respond with a failure response (or no response at all till timeout).
	\item This would enable the user to distinguish between System-level and Application-level errors, which can save lots of troubleshooting time in cases of numerous servers.
\end{itemize}

Provided these assumptions are met, the Server is ready to be setup and integrated into the code-base with just an import from the SerPyTor Package.

\subsection{Gateway}
\label{subsec:gateway}
The \textit{\textbf{Gateway}} object stores the context required for the associated Server objects, and stores the task routing information in them at regular intervals, or after the next task arrives - whichever comes first.
Tasks in a gateway are queued in either a single-level queue or a queue silo, depending on the task sorting and context determination algorithms. Appropriate sorting algorithms along with fallback mechanisms are provided, in case of failures to reduce the probability of a single point of failure, and to increase the probability of graceful degradation. This ensures that there are least surprises during real world, high-concurrency scenarios.

Similar to the Server Object, the Gateway object is weakly opinionated in nature to ensure maximum developer freedom while not compromising on overall setup integrity. It has a few minimal assumptions, which are summarized as below:

\begin{enumerate}
	\item A predefined \textbf{task to execute} (weak requirement).
	\item A \textbf{list of servers} setup along with their Heartbeat and Application Processes up.
	\item A predefined \textbf{allocation algorithm}, or a user-defined allocation algorithm that can handle all possible edge cases that may arise (please note that performance may take hit depending on implementation.)
\end{enumerate}

Ideally, the gateway object should possess only one task so to ensure the ideal scenario of durable execution (i.e., deterministic inputs and outputs). However, for more flexibility, users can configure them as custom tasks with a limited guarantee of durable execution. They can do it by mutating the task definition as and when required, subject to failure risks, and hence, not recommended unless there is no other way.

% \newpage
\section{Logical Execution Flow}
\label{sec:exec_flow}
\begin{figure}[!htb]
	\centering
	\includegraphics[width=16cm, height=8cm]{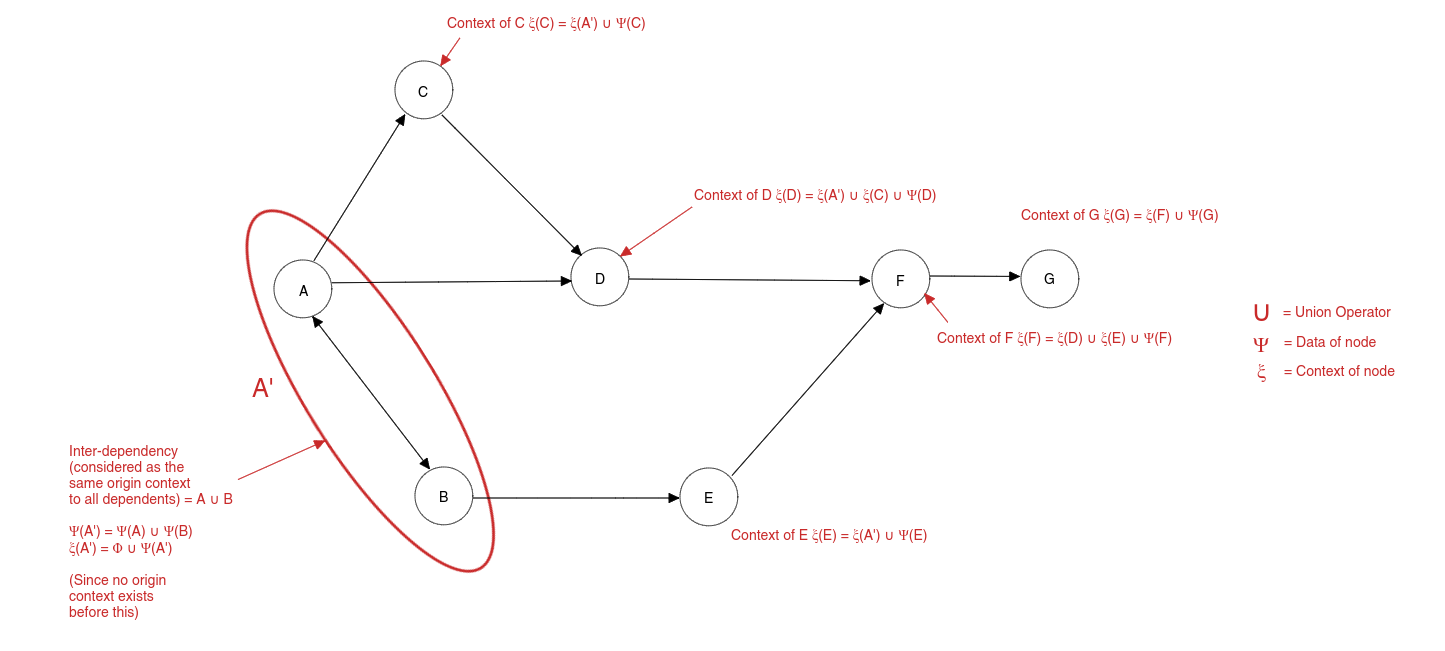}
	\caption{An example of logical execution flow in the distributed graph}
	\label{fig:node_exec}
\end{figure}

\subsection{Context Definition and Inheritance Mechanism}
Since the context is an integral part of the Graph Execution, it is imperative to define a robust system of mechanisms that involve context and its transference.

The fundamental unit of a computational graph is a node. In the case of context-aware graph, we cannot use the regular nodes of a graph. A context-aware graph can only be fulfilled if the nodes itself are context-aware, i.e., they possess context of some sort, which is relevant to their computation and output. To start defining the aforementioned context transference mechanism, we need to understand what can be the cases for each node.

A node's inherited context can be any of the following cases:

\begin{enumerate}
	\item Context from a Single Independent Origin
	\item Context from a Single Dependent Origin
	\item Context from Multiple Independent Origins
	\item Context from Multiple Dependent Origins
\end{enumerate}

We can ensure a robust mechanism of context transference using a set of simple rules:

\begin{enumerate}
	\item For the root node $\mathbb{R}$, it's context is defined as:
	      $$\xi(\mathbb{R}) = \xi(\mathbb{0}) \cup \Psi(\mathbb{R})$$, where:
	      \begin{itemize}
		      \item $\Psi(\mathbb{R})$ means the data of Node $\mathbb{R}$,
		      \item $\xi(\mathbb{0})$ means the origin context of the root node
	      \end{itemize}

	      The origin context of the root node is represented as $\xi(\mathbb{0})$, which can have the value of $\Phi$ or {} if there is no context prior to computation at the root node, i.e., no environment variables or similar data of the sort, supplied prior to the start of computation.
	      \\
	\item For a node with either \textbf{Single or Multiple independent origins}, its context is the union of the context of each origin node. In Figure \ref{fig:node_exec}, nodes G and F are respective examples.
	      \\
	\item For a node with either \textbf{Single or Multiple dependent origins}, its context is the union of the context of their parents and each node that is codependent on their parents, and other dependent nodes. In Figure \ref{fig:node_exec}, nodes A, B are dependent on each other and can be thought to have formed some kind of a \textbf{"union node"} \textit{A'} such that: $$\xi(A') = \xi(A) \cup \xi(B) \cup \Psi(A) \cup \Psi(B)$$. As seen in the figure, all children of A and/or B are transferred the origins of A' instead of A or B.

\end{enumerate}

\subsubsection{Possible Pitfalls}
\label{subsub:pitfalls}
A common possible pitfall of the aforementioned approach is that of the existence of possible loops in a graph. It is to be remembered here that the graph should be a \textbf{Directed Acyclic Graph}, commonly known as a DAG. This is a barebones necessity as even in code-bases that run on a single hardware resource (that is, a single system), dependencies cannot be circular in nature, and failure to do so gives rise to the \href{https://en.wikipedia.org/wiki/Circular_dependency#:~:text=are%20often%20encouraged.-,Problems,-%5Bedit%5D}{\textbf{Circular Import Problem}}, and \textit{Mohammed, Mawal, et. al.,} provide a search-based approach\cite{mohammed2022search} to minimize/eliminate circular dependencies . In extreme cases though, contextual circular dependency resolution techniques similar to what is mentioned by \textit{Athan, Tara, et. al.,}\cite{athan2013importation} can be applied.

\subsection{Durable Execution}
\label{subsec:durable_execution}
\textbf{Durable Programming} is a recent paradigm of context-aware programming that focuses on ensuring that tasks execute atomically. Instead of trying to maximize chances of successful execution by trying out edge cases, it aims to optimize for chances of failure by employing various methods to reduce failures. It does so by breaking down a callable entity into atomic units of computation that can be handled safely, and tractably - hence the name \textbf{Durable Execution}.

Platforms such as Temporal (or Azure Durable Functions) that closely follow the paradigm of durable execution, need deterministic outputs from workflows, which can be thought of as akin to deterministic outputs from the nodes of a graph. As such, context-aware nodes make a good use case of the durable programming paradigm according to the Vertex-centric Model described by \textit{Liu, et. al,}\cite{liu2020large}, with a small but significant change - the context needs to be propagated from origin during every call, as an individual context is deterministic in nature with a deterministic output, but collectively, they are non-deterministic in nature.

As mentioned in Section \ref{subsec:server}, Dependency Injection at run-time is the key to deterministic input. Making context-aware policies for dependency injection has been explored in other relevant literature, such as the one proposed by \textit{Ekstrand, Michael et. al}\cite{ekstrand2016dependency}.

\section{Conclusion and Future Works}
\label{sec:conclusion}

In this paper, we have introduced examined the potential of a new programming framework for distributed computations, employing context, context-aware policies and a modified form of durable programming, which is a radical shift from the traditional approach of programming. Scope for future work involves improving benchmarks by optimizing internal implementations of various algorithms, making troubleshooting easier through setting up guards at different execution checkpoints (that is., adapting a hybrid approach to maximize program durability as well as developer/user experience). In particular, implementation of the Gateway object and server response timings require a hefty amount of optimization, to extract maximum performance possible, so that task queuing and delegation bottlenecks are minimized. Otherwise, the bottlenecks will get magnified as soon as multiple Gateways are involved. Thus while scaling up, small bottlenecks might become an issue. Possible solutions involve rewriting IO and processing heavy tasks into compiled and memory-safe languages, such as Rust or D. Another alternative with possibly memory tradeoffs (insignificant in large-scale scenarios) would be resorting to functional languages like Scala and Erlang to speed up processes through real-time concurrency, along with Natively Implemented Functions (NIFs) written in compiled Object-Oriented Languages plugged in wherever necessary.

% \section*{Acknowledgments}
% This was was supported in part by......

%Bibliography
\bibliographystyle{unsrt}
\bibliography{references}

\end{document}